\documentclass[traditabstract]{aa}
\usepackage{graphicx}
\usepackage{txfonts}
\usepackage{natbib}

\newcommand{\sdss}{\mbox{SDSS1212--0123}}
\newcommand{\kmps}{km\,s$^{-1}$}
\newcommand{\msun}{M$_\odot$}
\newcommand{\rsun}{R$_\odot$}
\newcommand{\Mwd}{\mbox{$M_\mathrm{wd}$}}
\newcommand{\Msec}{\mbox{$M_\mathrm{sec}$}}
\newcommand{\Rwd}{\mbox{$R_\mathrm{wd}$}}
\newcommand{\dwd}{\mbox{$d_\mathrm{wd}$}}
\newcommand{\Rsec}{\mbox{$R_\mathrm{sec}$}}
\newcommand{\dsec}{\mbox{$d_\mathrm{sec}$}}
\newcommand{\Porb}{\mbox{$P_\mathrm{orb}$}}
\newcommand{\Teff}{\mbox{$T_\mathrm{eff}$}}
\newcommand{\logg}{\mbox{$\log g$}}

\newcommand{\Twd}{\mbox{$T_{\mathrm{eff,WD}}$}}                                
\newcommand{\Tsec}{\mbox{$T_{\mathrm{eff,sec}}$}}

\begin{document}

\titlerunning{Post common envelope binaries from SDSS. IV: \sdss, a new eclipsing system}

   \title{Post common envelope binaries from SDSS. IV: SDSSJ121258.25--012310.1, a new eclipsing system\thanks{This paper includes data gathered with the 6.5 meter Magellan Telescopes located at Las Campanas Observatory, Chile.}}

   \author{A. Nebot G\'omez-Mor\'an\inst{1}
     \and A.D. Schwope\inst{1}
     \and M.R. Schreiber\inst{2}
     \and B.T. G\"ansicke\inst{3}
     \and S. Pyrzas\inst{3}
     \and R. Schwarz\inst{1}
     \and J. Southworth\inst{3}
     \and J. Kohnert\inst{1}
     \and J. Vogel\inst{1}
     \and M. Krumpe\inst{1}
     \and P. Rodr\'iguez-Gil\inst{4}
   }

   \institute{Astrophysikalisches Institut Potsdam,
                An der Sternwarte 16, D-14482 Potsdam
              \and Departamento de Fisica y Astronomia, Universidad de
                Valparaiso , Avenida Gran Bretana 1111, Valparaiso, Chile
              \and
              Department of Physics, University of Warwick, Coventry, CV4 7AL,
                UK
              \and 
              Instituto de Astrof\'isica de Canarias, V\'ia L\'actea, s/n, La
                Laguna, E-38205, Tenerife, Spain \\
              \email{angm@aip.de}
   }
          \date{}       
\abstract{From optical photometry we show that SDSSJ121258.25--012310.1 is a new eclipsing, post common--envelope binary 
with an orbital period of $8.06$ hours and an eclipse length of $23$ minutes. We observed the object 
over 11 nights in different bands and determined the ephemeris of the eclipse to 
HJD$_\mathrm{mid}$ = $2454104.7086(2) + 0.3358706(5) \times \mathrm{E}$, where numbers in parenthesis indicate the 
uncertainties 
in the last digit. The depth of the eclipse is $2.85 \pm 0.17$ mag in the $V$ band, 
$1.82 \pm 0.08$ mag in the $R$ band and $0.52 \pm 0.02$ mag in the $I$
band. From spectroscopic observations we measured the 
semi-amplitude of the radial velocity $K_2 = 181 \pm 3$ km/s for the secondary star.
The stellar and binary parameters of the system were constrained from a) fitting the SDSS 
composite spectrum of the binary, b) using a $K$-band luminosty-mass relation 
for the secondary star, and c) from detailed analyses of the eclipse light
curve. The white dwarf has an 
effective temperature of $17700 \pm 300$ K, and its surface gravity is $\logg =7.53 \pm 0.2$. 
We estimate that the spectral type of the red dwarf is $\mathrm{M4} \pm 1$ and the 
distance to the system is $230 \pm 20$ parsec. The mass of the secondary star
is estimated to be in the 
range $\Msec = 0.26-0.29$ \msun\,, while the mass of the white dwarf is most
likely  $\Mwd = 0.46-0.48$ \msun. From an empirical mass-radius 
relation we estimate the radius of the red dwarf to be in the range
$0.28-0.31$ \rsun\,, whereas we get $\Rwd = 0.016-0.018$ \rsun\, from a theoretical mass-radius realation. 
Finally we discuss the spectral energy distribution and the likely
evolutionary state of \sdss.   

}
\maketitle

\section{Introduction}
\label{intro}
Most stars are found in binary or multiple star systems and a large
fraction of binaries will interact at some point in their lives. Interaction
depends on initial separation, relative masses of the components, and their
evolutionary state. When the initial orbital period is less
than $\sim5$ days, interaction will take place while both stars are on the main
sequence, giving rise to contact binaries and eventualy mergers. For wider
orbits, interaction can happen only when the more massive star reaches the RGB 
or the AGB. If the mass transfer rate is high enough a gaseous
envelope may surround the binary, entering thus a common envelope (CE)
phase. The stars spiral towards each other, friction within the CE will
lead to a shrinkage of the binary separation, and angular momentum and energy
are extracted from the orbit, expelling the CE. Once the envelope
is expelled, the remaining system consists of a main sequence star and a
remnant of the more massive star, e.g. a white dwarf, perhaps surrounded by
the ejected material that can be ionized forming a planetary nebula
\citep{Pac76,IbenLivio93}. Further evolution of the system is driven by
angular momentum loss due to magnetic braking and/or gravitational radiation,
which will bring the system into a semidetached configuration, becoming a
cataclismic variable (CV). Post common envelope binaries (PCEBs) are thought to be precursors of other interesting objects such as CVs, millisecond pulsars,
low-mass X-ray binaries or double degenerate white dwarfs. 

Though theories for the CE phase exist it is still poorly
understood. The efficiency of mass ejection in the CE with respect to the
masses of the components, their evolutionary state and the orbital separation
is uncertain \citep{Politano2007,TaamRicker06}. White dwarf plus main sequence
(WDMS) binaries are a perfect test laboratory for studying the current
population synthesis models because they are numerous, their stellar components 
are well understood in terms of their
single evolution and they are not accreting, which would increase the complexity. The number of PCEBs with well-defined parameters is still
small, although it has increased since \cite{Schr03} from 30 to almost 50 systems
\citep{2004A&A...418..265G,2005MNRAS.359..648M,2006A&A...456.1069S,2007A&A...466.1031V,2007A&A...469..297A,2007A&A...474..205T,Schreiber2008,Rebassa2008,Steinfadt2008}.
 
In this paper we report the discovery of a new eclipsing PCEB. 
In our ongoing search for PCEBs among white-dwarf/main-sequence binaries
\citep{Schreiber2008,Rebassa2007,Rebassa2008}, SDSSJ121258.25--012310.1 \citep{2008ApJS..175..297A} (henceforth \sdss) was
included in our target list for photometric monitoring of candidate 
objects. The serendipituous discovery of a binary eclipse from time-resolved
differential photometry triggered a photometric and spectroscopic
follow-up. Only seven eclipsing binaries containing a white
dwarf and a low mass main sequence star were known until 2007. Since then another three 
eclipsing systems have been published \citep{Steinfadt2008,Drake2008}, and a further three systems have been discovered by us \citep{Pyrzas2008}. Eclipsing binaries are of great interest since they offer the possibility of deriving fundamental properties of stars with a high accuracy. \sdss\ was firstly listed as a quasar candidate by \cite{Richards2004} and later classified as a $\mathrm{DA+dMe}$ by \cite{Silvestri2006}. It contains a relatively hot white dwarf (from now on primary) and an active mid-type dM star (from now on secondary).

In this paper we summarize our current knowledge about this source from own
observations and archival work. It is organized as follows. In Sect.\,\ref{Obser} we describe the observations and reductions. In
Sect.\,\ref{resul} we present the results, we study the evolution of the system
in Sect.\,\ref{evolu} and conclude in Sect.\,\ref{disc}. 

\section{Observations and reductions}
\label{Obser}
\subsection{IAC80 and AIP70 photometry}
\label{phot}
\begin{figure}[t!]\centering
\includegraphics[width=8cm]{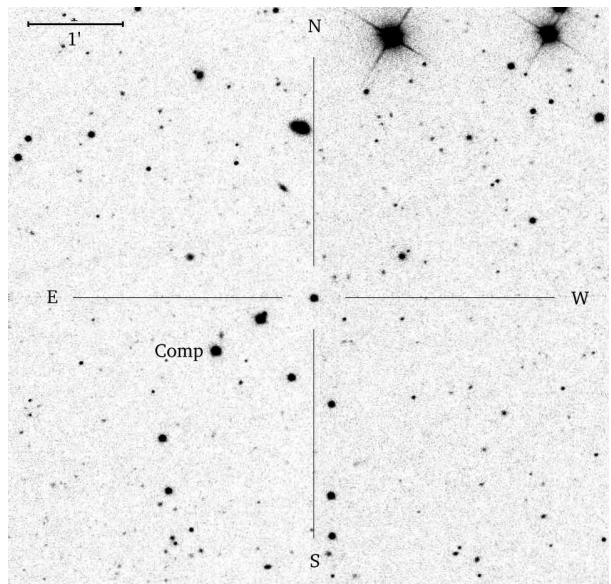}
\caption{SDSS image of \sdss\, (in the cross-hair) and the
  comparison star ($RA=12$:$13$:$02.39$, $DEC=-01$:$23$:$43.5$). } 
\label{g:finding_chart}     
\end{figure}

Optical photometric observations were obtained using two different telescopes over
 11 nights. The 80 cm telescope IAC80 in Observatorio del Teide,
Spain, was equipped with the standard CCD camera and the 70 cm telescope
of the Astrophysical Institute Potsdam at Babelsberg was used with a
cryogenically cooled 1x1 k TEK-CCD. A log of observations is presented in
Table~\ref{tab1}. A field of $\sim3$ arc minutes was read with the IAC80 CCD camera,
and we used a binning factor of 2 in both spatial directions (scale of
0.6''), while we used a binning factor of 3 for the 70 cm telescope (scale of
1.41"), in order to decrease the readout time and improve the signal to 
noise. Reduction was performed using standard packages in IRAF\footnote{IRAF is distributed by the National Optical Astronomy Observatory, which is operated by the Association of Universities for Research in Astronomy, Inc., under contract with the National Science Foundation, http://iraf.noao.edu} and MIDAS. Differential magnitudes were obtained with respect to the comparison star SDSS J121302.39--012343.5 (see Fig.\ref{g:finding_chart}), with magnitudes {\em ugriz}=$17.40,16.00,15.51,15.36,15.30$. SDSS magnitudes were transformed into Johnson's using equations taken from the Sloan pages\footnote{http://www.sdss.org/dr7/algorithms/sdssUBVRITransform.html}. Neglecting the color term, we calculated absolute magnitudes of \sdss. The estimated error of the absolute calibration is $0.05$ mag.  

\begin{table}[h!]
\caption{Log of photometric observations for \sdss. }
%IAC80 is the $80$
%  cm telescope of in Observatorio del
%Teide and AIP70 is the $70$ cm telescope at Babelsberg. $\mathrm{Tel}$ is the telescope,
%$t_\mathrm{{int}}$ is the integration time in seconds, ${N_\mathrm{obs}}$ stands for the number
%of exposures taken, $\phi_\mathrm{{ini}}$ and $\phi_\mathrm{{fin}}$ are the phase coverage.}
\centering
\label{tab1}       
\begin{tabular}{rrrrrrr}
\hline\hline 
$\mathrm{Date}$ & $\mathrm{Tel}$ & $\mathrm{Filter}$ & $t_\mathrm{{int}}$  & $N_\mathrm{{obs}}$ & $\phi_\mathrm{{ini}}$ & $\phi_\mathrm{{fin}}$\\[3pt]
\hline\noalign{\smallskip}
04 Jan 2007 & IAC80 & $I$ & 180 & 74  & 0.805 & 1.278 \\
26 Jan 2007 & AIP70 & $V$ & 180 & 19  & 0.041 & 0.156 \\
13 Feb 2007 & IAC80 & $V$ & 70  & 191 & 0.607 & 1.202 \\
14 Feb 2007 & IAC80 & $V$ & 70  & 221 & 0.525 & 1.292 \\
12 Mar 2007 & AIP70 & $I$ & 120 & 181 & 0.031 & 0.946 \\
13 Mar 2007 & AIP70 & $I$ & 120 & 153 & 0.377 & 1.343 \\
14 Mar 2007 & AIP70 & $I$ & 120 & 49  & 0.786 & 1.040 \\
15 Mar 2007 & AIP70 & $R$ & 120 & 73  & 0.683 & 1.023 \\
26 Mar 2007 & AIP70 & $I$ & 120 & 45  & 0.994 & 1.179 \\
21 May 2007 & AIP70 & $I$ & 90  & 59  & 0.907 & 1.097 \\
06 May 2008 & IAC80 & $R$ & 120 & 32  & 0.913 & 1.049 \\
\noalign{\smallskip}\hline
\end{tabular}
\end{table}

\subsection{Spectroscopy}
\label{spec}
Spectroscopic follow up observations were obtained during the period 16-19 May
2007 with the LDSS3 imaging
spectrophotograph at the Magellan Clay telescope. Ten spectra were taken for
\sdss. Exposure times varied from 300 to 600 seconds. Seeing and
transparency were highly variable. The VPH\_Red grism and an OG590 blocking
filter were used. The detector was a STA 4k$\times$4k pixel CCD with two read out
amplifiers. We used a slit width of 0.75 arcsec, that together
with the spectral resolution $R=1810$, gave a coverage of
$5800-9980$ \AA\ at a reciprocal dispersion of $1.2$ \AA pix$^{-1}$. Four of the
spectra taken at quadrature were obtained through a narrow slit of $0.5$
arcsec resulting in a FWHM spectral resolution of $4.8$\,\AA, with the purpose
of measuring the radial velocity amplitude with a higher accuracy.  
Flat-field images were taken at the position of the target to allow 
effective fringe removal in the red part of the spectra.
The spectral images were reduced using STARLINK packages FIGARO and KAPPA, and the spectra were
optimally extracted \citep{1986PASP...98..609H} using the PAMELA package
\citep{1989PASP..101.1032M}. Wavelength calibration was done 
using sky lines. Wavelengths of good sky lines were obtained from the atlas of
\cite{1996PASP..108..277O,1997PASP..109..614O}. A fifth-order polynomial was fitted to 
$36$ sky lines. Spectra were flux calibrated and corrected for telluric
lines using spectra of the standard star LTT3218 taken during the same observing run. 

\section{Results}
\label{resul}

\subsection{The light curve}
\begin{figure} [t!]
\centering
\includegraphics[width=10cm,angle=-90,clip=]{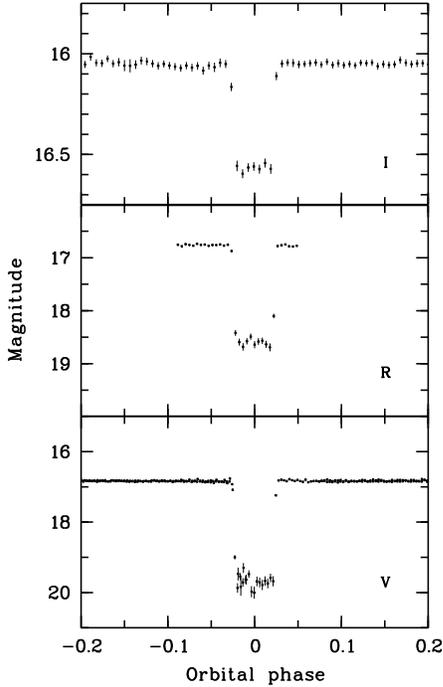} 
\caption{Optical photometry from the IAC80 telescope in the $V$, $R$ and $I$ band (from bottom to top) phase folded over the orbital period. Note the different scales for each panel.}\label{g:eclipse}
\end{figure}

The optical light curve of \sdss\, displays a total eclipse of the primary
with length of approximately $23$\, minutes. The depth of the eclipse is $0.52
\pm 0.02$ mag in the $I$ band, $ 1.82 \pm 0.08$ mag in the $R$ band and $2.85 \pm
0.17$ mag in the $V$ band (see Fig. \ref{g:eclipse}). Eclipse magnitudes are
$m_I = 16\fm56\pm0\fm02$, $m_R = 18\fm58\pm0\fm08$ and $m_V =
19\fm68\pm0\fm17$. The much deeper eclipse in 
the $V$ band is due to the fact that the primary emits most of the light in the
blue, while the secondary dominates in the $I$ band. Photometric variability
outside of the eclipse, e.g. from an irradiated secondary or from ellipsoidal
modulation of the secondary, was found to be less than $0\fm01$. 

At the given time resolution of our photometry, the WD ingress and egress phases are
not resolved. Five eclipses were completely covered and the eclipse length was
determined in these light curves measuring their full width at half
maximum of the flux level. The weighted mean of those five measurements gives an eclipse length
of $23 \pm 1$\,min.

\subsection{Ephemeris}
In addition to the five eclipses which were covered completely one further
eclipse was covered partially. Using the measured eclipse length from the
previous section we thus determined six eclipse epochs
(Table~\ref{t:eclipse}). The eclipses of March 12, 14 and 26, respectively,
were not covered due to bad weather conditions.  
Using a phase-dispersion minimization technique a tentative period was
determined, $\Porb = 0.3359 \pm 0.0006$\,hours, which was sufficiently
accurate to connect all follow-up observations without a cycle count alias. 

We then used the six mid eclipse epochs to calculate a linear ephemeris by
fitting a line to the cycle number and eclipse epoch:  
\begin{equation}
\rm{HJD}_{mid} = 2454104.7086(2) + 0.3358706(5) \times \mathrm{E},
\end{equation} 
where numbers in parenthesis indicate the $1\sigma$ uncertainty in the last
digits. The observed minus calculated values are tabulated in Table~\ref{t:eclipse}.  

\begin{table}[h!]
\centering
\caption{Date, times of mid eclipses, cycle number obtained from the
  photometric observations %carried out with the IAC80 and the AIP70cm telescopes and
  and residuals from the linear ephemeris. }%Errors of mid-eclipse times are shown
%  in parenthesis.}
\label{t:eclipse}
\begin{tabular}{rrrr}
\hline\hline 
$\mathrm{Date}$ & $\mathrm{HJD}$ (Mid-eclipse) & $\mathrm{Cycle}$  & $O-C$ (s)\\[3pt]
\hline\noalign{\smallskip}
04 Jan 2007$^c$ & {2454104}{.}{7085(21)} & 0 & -9.9\\
13 Feb 2007$^c$ & {2454145}{.}{6847(8)} & 122 & -11.0\\
14 Feb 2007$^p$ & {2454146}{.}{6922(8)} & 125 & -19.0\\
13 Mar 2007$^c$ & {2454173}{.}{5621(10)} & 205 & -0.3\\
21 May 2007$^c$ & {2454242}{.}{4163(21)} & 410 & 66.9\\
06 May 2008$^c$ & {2454593}{.}{4000(14)} & 1455 & -25.9\\
\noalign{\smallskip}\hline
\end{tabular}

$^c$ Eclipse completely covered.\\
$^p$ Eclipse partially covered.
\end{table}

\subsection{Stellar parameters}
\label{parameters}
\subsubsection{Decomposition of the SDSS spectrum}
\label{fitting}
We determined the stellar parameters of \sdss\, from the SDSS spectrum
following the procedure described in \cite{Rebassa2007}. 

\begin{figure}[t!]
\centering
\includegraphics[width=6.5cm,angle=-90]{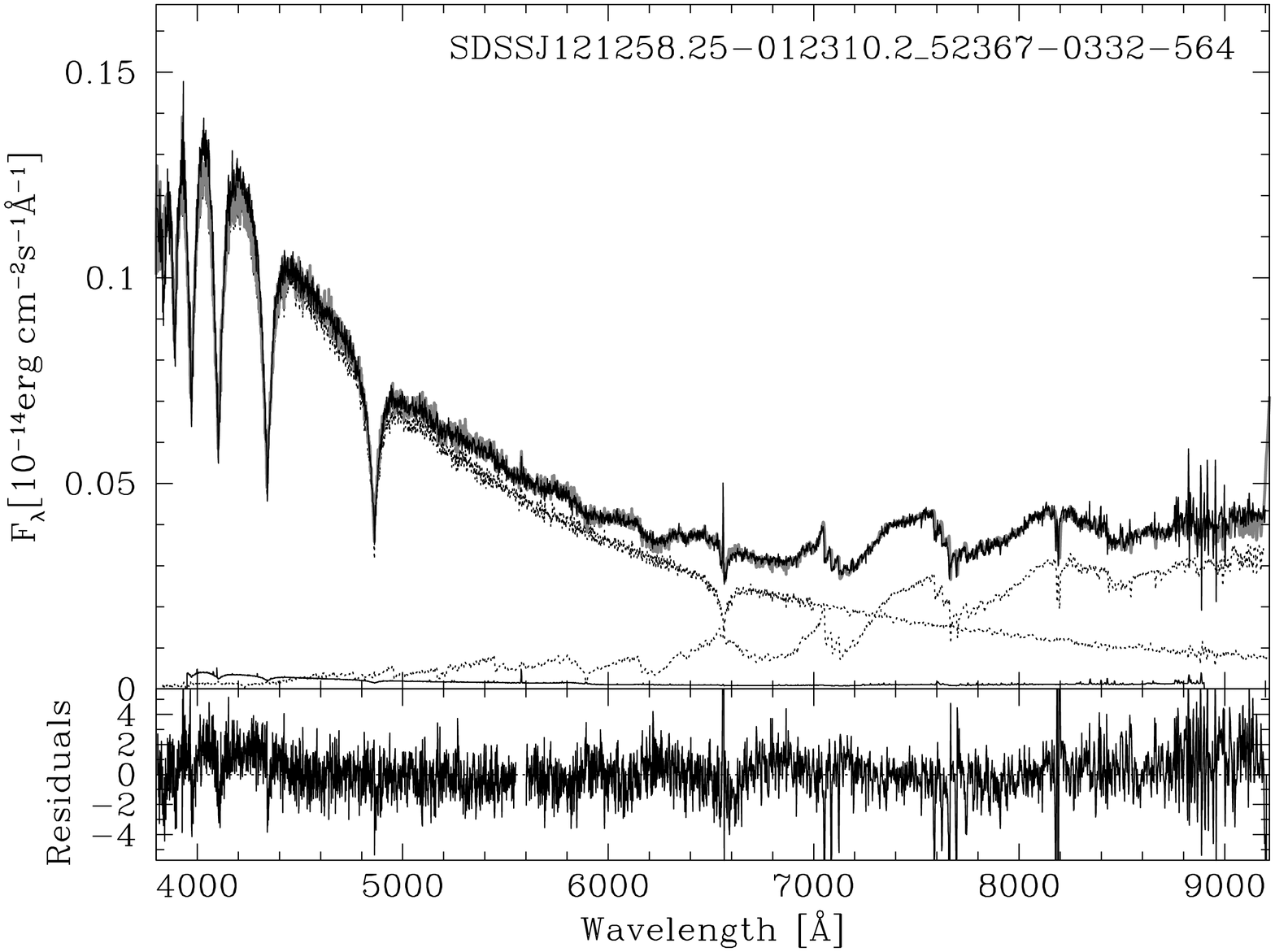}
\caption{Two component fit to \sdss. The top panel shows the WDMS spectrum
  (black line) and the white dwarf and the $\mathrm{M4}$ M-dwarf templates (dotted lines),
  while the lower panel shows the residuals to the fit.}
\includegraphics[width=8.5cm]{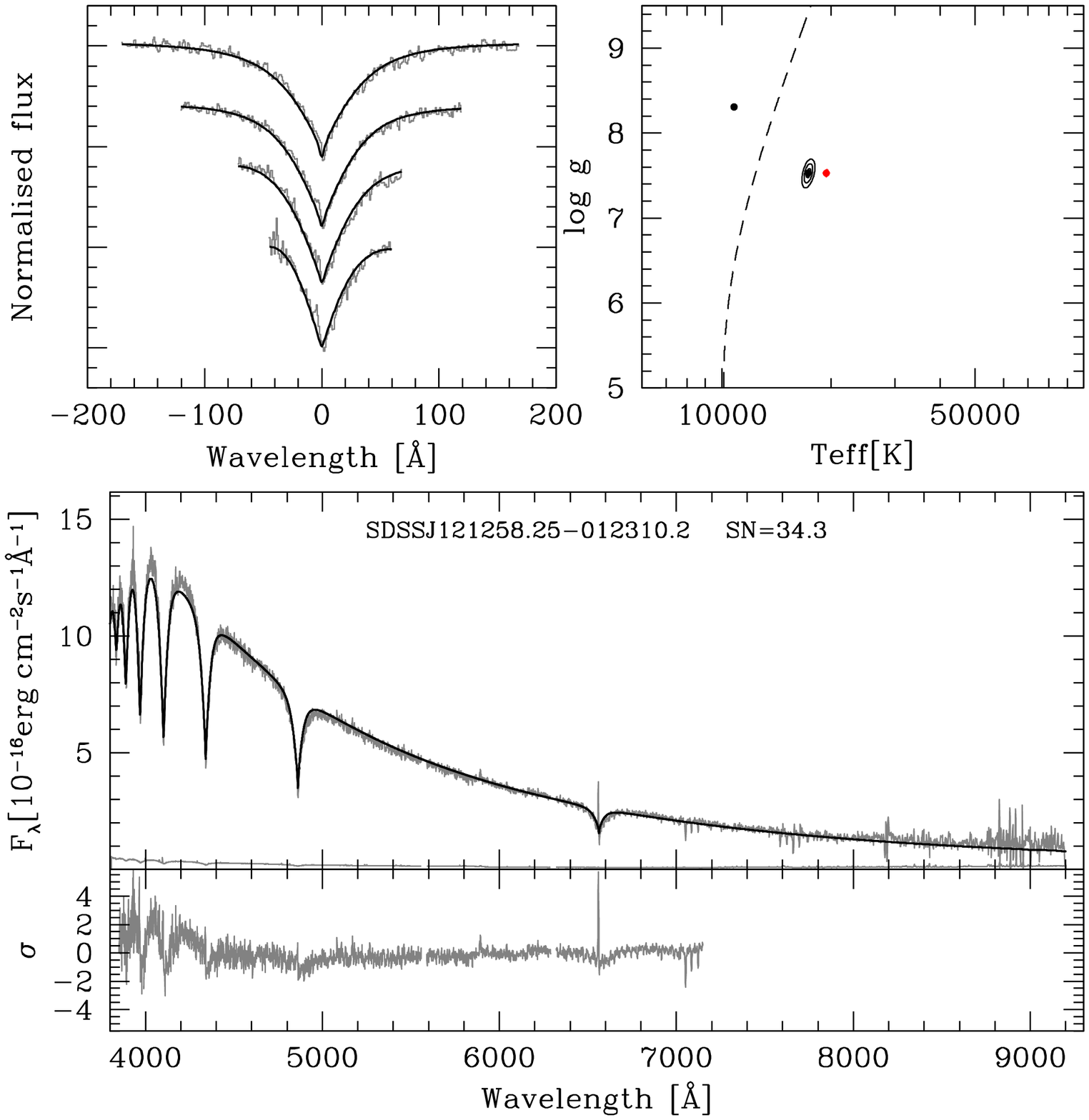}
\caption{Spectral model fit to the white dwarf component of \sdss,
  obtained after subtracting the best-fit M-dwarf template from its SDSS
  spectra. Top left panel: best fit (black lines) to the normalized H$\beta$
  to H$\epsilon$ line profiles (gray lines, top to bottom). Top right panel:
  $1$, $2$ and $3\sigma$ $\chi^{2}$ contour plots in the $T_{\rm eff}-log\rm{g}$
  plane. The black contours refer to the best line profile fit, the red
  contours to the fit of the whole spectrum. The dashed line indicates where
  the maxima of the H$\beta$ equivalent width occurs in the $\Teff$--$\logg$\, plane, dividing it into two different solutions, a cold and
  a hot one. The best-fit parameters of the hot and the cold normalized line
  profile solutions and of the fit to the $3850$ -- $7150$ \AA\,  range are indicated
  by the black and the red dots, respectively. Bottom panel: the white dwarf
  spectrum and associated flux errors (gray lines) along with the best-fit
  white dwarf model (black lines) to the $3850$ -- $7150$ \AA\, wavelength range
  (top) and the residuals of the fit (gray line, bottom).} 
\label{g:param}
\end{figure}

In a first step the best match of the SDSS composite spectrum is determined 
with an optimization strategy on a grid of observed white dwarf and M-dwarf
template spectra created from the SDSS DR6 database. The main result of this first
step is the determination of the spectral type of the secondary. 
Using the spectral type-radius relation from \cite{Rebassa2007} and
the apparent magnitude of the scaled template results in a first distance
estimate $\dsec$. 
After subtracting the best-fitting M-star template, white-dwarf parameters are
determined via $\chi^2$ minimization in a $\logg$ -- $\Teff$ grid of model
atmospheres \citep{2005A&A...439..317K}. Since this analysis step is performed
on spectra normalized to a continuum intensity, the results are bi-valued
yielding a `hot' and a `cold' solution (see Fig.~\ref{g:param}). 
The degeneracy can typically be broken by an additional fit to 
the overall spectrum (continuum plus lines in the wavelength range
$3850$ -- $7150$\,\AA). In the present case of \sdss\, the GALEX detection
(see below) provides an additional constraint excluding the `cold' solution.
The results of the spectral decomposition and the white dwarf fit for \sdss\,
are shown in Fig.~\ref{g:param}. 

Mass and radius of the white dwarf are calculated with the best-fitting
$\log g$ -- $\Teff$ combination using updated versions of the tables by
\cite{1995ApJ...449..258B}.
The flux scaling factor together with the derived radius of
the white dwarf results in a second distance estimate of the binary, $\dwd$.

The spectral type of the secondary was determined to be $\mathrm{M4}\pm1$ implying a
distance $\dsec=320 \pm 95$ pc, mass range of the secondary $\Msec = 0.255$ -- $0.380$ \msun\, 
and radius range $\Rsec = 0.258$ -- $0.391 $ \rsun,  using \cite{Rebassa2007} spectral 
type-mass and spectral type-radius empirical relations respectively. The derived temperature and 
$\log g$ of the primary were found to be only weakly dependent on the chosen spectral 
type and spectral template of the secondary, because we use H$\beta$ -- H$\epsilon$ for
the white dwarf line fit, where the secondary star contribution is small. It is also weakly dependendt on
 the accuracy of the spectral flux calibration and also the small radial velocity line displacements.
The best fit was found for $\Teff=17700 \pm 300$ K and $\log g=7.53 \pm
0.05$ (implying a white dwarf mass $\Mwd = 0.39 \pm 0.02$\,\msun, and $\Rwd =
0.018 \pm 0.001$\, \rsun). The obtained values are in agreement 
with those published by \cite{Silvestri2006}.  
However, one should be aware of the fact that all the quoted 
errors are purely statistical. The true uncertainty of the white dwarf 
spectral parameters is clearly higher than suggested by the derived numbers. 
We estimate the systematic
uncertainty of our $\log g$ determination to be on the order of $0.2$ dex, which
results in rather wide ranges of possible values for the mass and 
the radius of the primary, i.e. 
$\Mwd = 0.33$ -- $0.48$ \msun\, and $\Rwd = 0.015$ -- $0.021$ \rsun.

The derived distance to the white dwarf is $\dwd = 226 \pm 8$\,pc (assuming
the statistical error only). 
The two distance estimates differ, $\dsec$ being longer than
$\dwd$, but in agreement within the erros. \cite{Rebassa2007} found a similar 
trend for $101$ WDMS binaries in their study. They argue 
that such difference could be due to stellar activity of the
secondary star, and that the spectral type determined from the optical SDSS 
spectrum is too early for the mass of the secondary star, which would lead to 
a larger radius and consequently a larger distance to the system. Since the
secondary in \sdss\, was found to be active too, we regard the distance
estimate for the white dwarf being more reliable. Taking into account 
systematic errors we obtain $\dwd = 230 \pm 20$\,pc as the distance 
to the system.

\subsubsection{Constraining the secondary mass using 2MASS}
\label{sec-2mas}

In the previous section we derived the mass and the radius of the 
secondary star using empirical relations from \cite{Rebassa2007} 
and obtained $\Msec = 0.255-0.380$ \msun\, and $\Rsec = 0.258-0.391$ \rsun,
respectively. However, as clearly shown in Fig.\,7 of  \cite{Rebassa2007}, 
the masses and radii derived from observations largely scatter around the
empirical relations. In addition, according to
\cite{Rebassa2007} increased activity of the rapidly rotating secondary stars in close binaries
can cause the stars to appear as earlier spectral types when compared to 
non-active stars of the same mass. To sum up, the secondary masses derived from
empirical relations can obviously only considered to reasonable but rough
estimates. 

An alternative method to determine the mass of secondary star is to use
luminosity-spectral type relations. To that end, we explored the Two Micron 
All Sky Survey Point Source Catalog \citep{2003tmc..book.....C}, finding 
magnitudes $J=14.90 \pm 0.03$, $H=14.39 \pm 0.05$ and $K_\mathrm{s}=13.96 \pm 0.05$ 
for \sdss. Subtracting the extrapolated contribution of the primary star 
($\log g = 7.5$ and $\dwd=230$ pc) yields infra-red colors 
of $J-H=0.51\pm 0.06, H-K_s=0.43\pm 0.07$, respectively.  
Using the empirical mass-luminosity relation from
\cite{Delfosse2000}, we derive the mass of the secondary star to be
$0.26 \pm 0.03$. Using again the mass-radius relation 
from \cite{Rebassa2007} this implies a spectral type $\mathrm{M5}$, i.e. 
later by one stectral type than estimated from the deconvolution 
of the SDSS spectrum. This supports the 
idea of activity significantly affecting 
the determination the secondary star spectral types and the corresponding 
distances.

\subsubsection{Radial velocity}
\label{period_rv}
In each of our observed spectra we measured the radial velocities of the NaI
absorption doublet ($8183.27$\AA, $8194.81$\AA), which originates 
from the secondary star. A double Gaussian with a fixed
separation of $11.54$ \AA\, was fitted to the line profiles using the FIT/TABLE
command provided by ESO/MIDAS. 

H$\alpha$ was deconvolved into an absorption and an emission line component
using two Gaussians.
While the emission line showed pronounced wavelength shifts, the centroids of
the absorption lines thus measured did not constrain the curve
of the white dwarf significantly.

\begin{figure}[t!]
\centering
\includegraphics[width=8.5cm,height=5.5cm,angle=0]{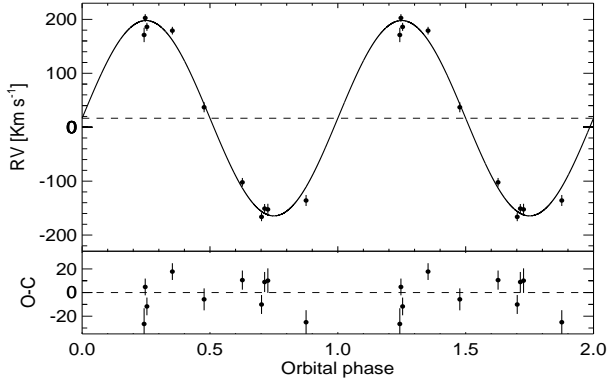}
\caption{Radial velocities measured from the NaI doublet $8183,8194$\,\AA\, originating from the
  secondary star of \sdss\, folded over the orbital period obtained
  from the photometry. Sine fit and residuals (lower panel) are shown.}
\label{g:rv_fig}
\end{figure}

Assuming a circular orbit a sine-function was fitted to the measured radial
velocity curves to obtain the radial velocity semi-amplitude $K_2$ of the
secondary star:
\begin{equation}
v_r = \gamma_2 + K_2 \sin\left[\frac{2\pi(t-t_\mathrm{0})}{P}\right],
\end{equation}
The orbital period $P$ and the epoch of mid eclipse $t_\mathrm{0}$ were determined
photometrically and were kept fixed for the radial velocity fit. 
For the NaI doublet we find the systemic velocity $\gamma_2= 17 \pm 3$\,\kmps\ 
and \mbox{$K_2 = 181 \pm 3$}\,\kmps , while we find for the H$\alpha$ line
$\gamma_2= 21 \pm 2$\,\kmps and $K_2 = 161 \pm 3$\,\kmps. 
The fit to the NaI lines is shown if Fig.~\ref{g:rv_fig} together with the
residuals. 

The semi-amplitudes of the two radial velocity curves are different
and these differences seem to be significant. The semi-amplitude derived 
from H$\alpha$ is lower, indicating that its emission
is displaced towards the inner hemisphere of the secondary star with respect
to the NaI doublet. As neither of the two line features
shows significant photometric variability, which would indicate a biased 
origin of one of the line species (e.g.~towards the non-irradiated side of 
the secondary), we exclude irradiation as the explanation for the
observed difference in $K_2$. A detailed comparison of radial velocities derived from the NaI doublet and H$\alpha$ lines has been 
performed by \citet{Rebassa2007}. They find that both velocities
often significantly differ but that there seems to be no systematic shift of 
H$\alpha$ radial velocities towards smaller values. As discussed in detail in 
\citet{Rebassa2008}, this is probably explained by the H$\alpha$ 
emission being related to activity and not uniformely distributed 
over the surface of the secondary. \cite{Kafka2005} studied in detail the origin of different line species, however \sdss1212\, shows no evidence of accretion nor irradiation. We therefore assume that in \sdss\, the NaI doublet much better traces the center of mass of the secondary and we use its semi-amplitude for the mass estimate.

\begin{figure} [t!]
\centering
\includegraphics[width=7.5cm,angle=0,clip=]{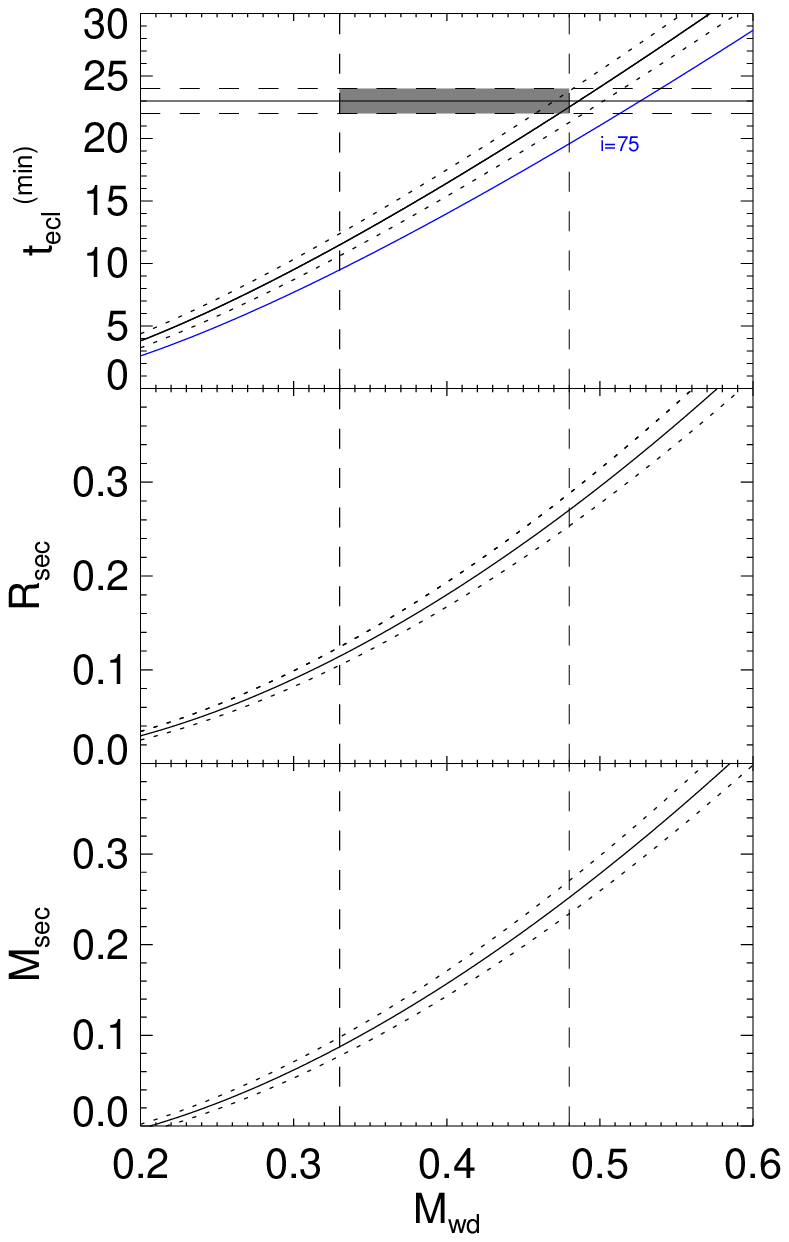}
\caption{ Solution of the mass function for $K_2=181\pm3$ km/s as a function of
$\Mwd$ (bottom panel); radius from the mass-radius empirical relation from \cite{Bayless2006} (mid panel); eclipse duration (top panel), assuming 
circular orbit and an inclination angle of $90{\degr}$. Vertical lines indicate
 the mass of the primary star, $\Mwd = (0.33-0.48)$ \msun, as determined from 
the deconvolution of the SDSS spectrum. The eclipse length, $t_\mathrm{ecl} = 23\pm1$ 
min, is marked with horizontal lines, and the intersection in shaded in grey. 
The erros in $K_2$ are shown with the small dashed lines in the three panels. 
In the top panel the solution for an inclination of $75{\degr}$ is plotted with 
a blue line to show it's influence. See text for a more detailed description.} 
\label{g:mass_rad}
\end{figure}

We write the mass function of the binary assuming a circular orbit in the form
\begin{equation}
\Msec = \left(\sqrt{\frac{2\pi G\sin^3i}{PK_2^3}\Mwd}-1\right)\Mwd,
\label{massfunc}
\end{equation}
and derive an upper limit for $\Msec$ 
for a given white-dwarf mass $\Mwd$ assuming $i=90\degr$ 
(see bottom panel of Fig.~\ref{g:mass_rad}).

Using the empirical mass-radius relation for main sequence stars derived by
\cite{Bayless2006} we estimate the radius of the secondary (middle
panel of Fig.~\ref{g:mass_rad}). The top panel of the same figure illustrates
the maximum possible eclipse length ($i=90\degr$, black line) for the given
stellar radius, the orbital period $P$ and the orbital separation $a$
according to  
\begin{equation}
t_{\rm ecl} = \frac{\Rsec P}{\pi a}. 
\end{equation}

The measured values of the eclipse length and the range of the
white-dwarf mass from Sec.\,\ref{fitting} are shown in the figure with
horizontal and vertical lines, respectively, their intersection is shaded in
grey in the top panel. It is also plotted the solution for $i=75 \degr$ for comparison. 
From the eclipse length the range of
possible values for the mass of the WD is $\Mwd = 0.46-0.52$, and for the dM
$\Msec = 0.21-0.32$ \msun, and $\Rsec = 0.23-0.34$ \rsun.

\subsubsection{Light curve modeling}
\label{sec-lcf}
A determination of most of the physical parameters of an eclipsing system can
be achieved by fitting model light curves to the actual data. We made use of
a newly developed light curve fitting code, written by T.R. Marsh, for the general
case of binaries containing a white dwarf.

The code is described in detail in \cite{Pyrzas2008}. Briefly,
a model light curve is computed based on user-supplied initial system
parameters. These are the two radii, scaled by the binary separation, $\Rwd/a$
and $\Rsec/a$, the orbital inclination, $i$, the unirradiated stellar
temperatures 
of the white dwarf and the secondary star \Twd\, and \Tsec\, respectively, the
mass ratio $q = \Msec/\Mwd$ and $t_\mathrm{0}$ the time of mid-eclipse of the white
dwarf. 

Starting from this parameter set, the model light curve is then fitted to the
data using Levenberg-Marquardt minimisation. Every parameter can either be
allowed to vary or remain fixed, during the fitting process.

Our approach for modeling the $I$ band photometry of \sdss\, was the
following. A large  
and dense grid of points in the $\Mwd-\Msec$ plane was first calculated, 
generously
bracketing the estimates for the mass of the two components (see
Sec.\,\ref{period_rv}). 
Each point defines a mass ratio $q$, and through $\Porb$, a binary separation
$a$. 
Furthermore, from the mass function equation (Eq. \ref{massfunc}), using the
value of $K_{2}$ (derived in Sec.\,\ref{period_rv}) and $\Porb$, one can
calculate the inclination angle $i$. Points for which (formally) $\sin i > 1$
were discarded from the grid, for all other points the corresponding light
curve model was computed, leading to the computation of some 9000 models.

As an initial estimate for the radii of the binary components, we adopted 
values from the theoretical M -- R relations of Bergeron et al (1995) for the
white dwarf, and \cite{Baraffe1998} - the $5$ Gyr model - for the
secondary. Regarding the two temperatures, \Twd\, and \Tsec\,, the value from 
our spectral decomposition (Sec.\,\ref{fitting}) was used for the white dwarf,
while the $\mathrm{Sp(2)}$ -- $\mathrm{T}$\, relation from \cite{Rebassa2007}, together
with our result for the spectral type of the secondary, were used to obtain an
initial value for \Tsec. 

For the fitting process $q$, $i$, \Rwd\, and \Twd\, were fixed, leaving only
\Rsec, \Tsec\, and $t_\mathrm{0}$\, free to vary. \Rwd\, was fixed mainly because of
the poor temporal resolution of our data set, which does not resolve the white
dwarf ingress and egress. Consequently, if allowed to vary, the white dwarf
radius would only be loosely constrained and it would introduce large
uncertainties in the determination of \Rsec. \Twd\, was also fixed, because
allowing both temperatures to vary simultaneously would lead to a degenerate
situation, as they are strongly correlated. Our spectral decomposition results
are sufficiently accurate, so as to allow us to fix \Twd\, without affecting
the fitting result. The parameter $t_\mathrm{0}$\, on the other hand, was left free
during the fitting, to account for the $O-C$ errors in the mid-eclipse times,
which in some cases were significant (see Table~\ref{t:eclipse} again).

The results of the light curve fitting process were analyzed as follows.
We first applied a cut in the quality of the fits. This was done by selecting
the minimum $\chi^{2}$ value of all fits and then culling all model fits at 
$>1\sigma$ above the best fit. Afterwards, we selected from the
remaining, equally good light curve fits, those which where physically
plausible. We defined a $\delta R$ parameter, as
$\delta R = \left(R_{\mathrm{fit}}-R_{\mathrm{th}}\right)/R_{\mathrm{th}}$,
i.e. how much the fitted radius value deviates from the theoretical radius
value, obtained from a M-R relation, for a given model. Thus, we selected only
those models that had $\delta R \le 0.15$, to allow for an oversized secondary.

\begin{figure} [h!]
 \centering
 \includegraphics[width=8cm]{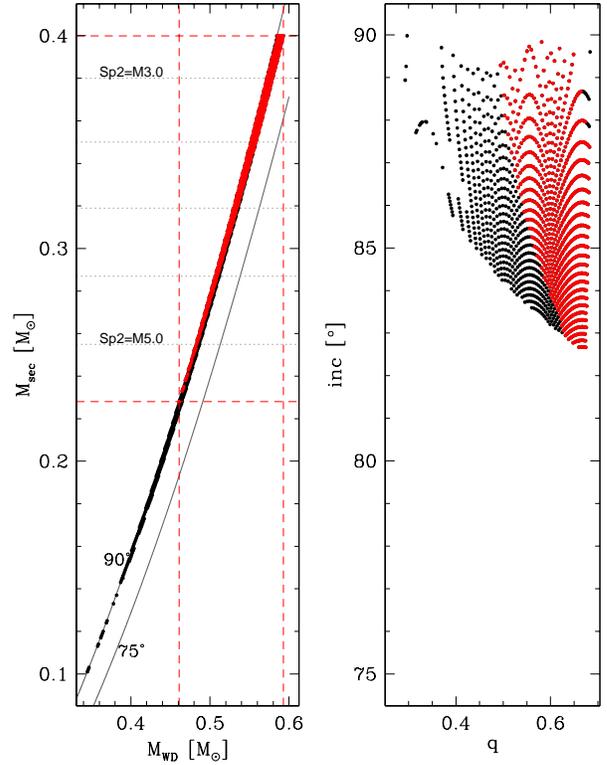}
  \caption{Light curve model fitting results for SDSS\,1212-0123. Left
    panel: \Mwd\, and \Msec\, values corresponding to fits with $\chi^{2}$
    values within $1\sigma$ of the minimum value (black points) and,
    simultaneously, with $\delta R \le 0.05$
    (red points). Right panel: the same, only in the $q-i$ plane. Also
    depicted in the left panel are curves corresponding to the mass
    function (solid black lines, $i=90^{\circ}$ and $75^{\circ}$)
    which (by definition) bracket the possible solutions,
    $\mathrm{Sp}(2)-\mathrm{M}$ relations (dotted, horizontal, gray
    lines) and the range of possible $\left(\Mwd,\Msec\right)$ values
    (dashed, horizontal and vertical, red lines)}
  \label{1212fitall}
\end{figure}

The results are illustrated in Figure\,\ref{1212fitall}. Black dots designate
those light curve fits making the 1$\sigma$ cut, red dots those that satisfy
both the 1$\sigma$ and $\delta R=15\%$ cuts. The resulting ranges in white
dwarf masses and secondary star masses (indicated with dashed, vertical, red
lines) are $\Mwd=0.46-0.6$ \msun\, and $\Msec=0.23-0.4$ \msun,
respectively, corresponding to a white dwarf radius of
$\Rwd=0.013-0.016$ \rsun\, and a secondary radius of
$\Rsec=0.27-0.41$ \rsun. The range for the inclination angle is
$i=82^{\circ}-90^{\circ}$. Also indicated, with dotted, horizontal, gray lines
are the radii of M-dwarfs with spectral types
$\mathrm{Sp}(2)=\mathrm{M}3-\mathrm{M}5$ in steps of 0.5, based on the
spectral type-mass relation given by \cite{Rebassa2007}. 

\begin{figure} [h!]
 \includegraphics[angle=-90,width=8cm]{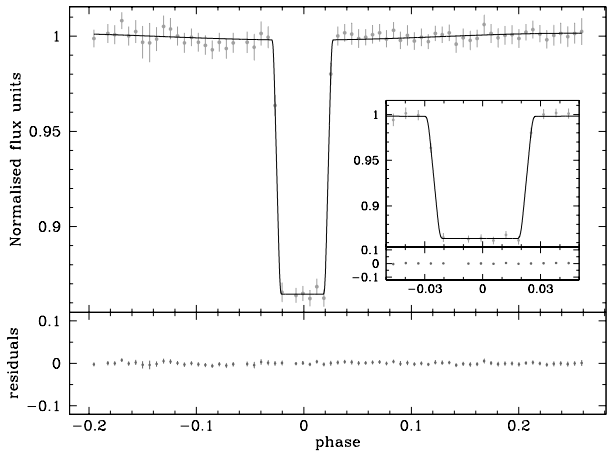}
  \caption{Model fit to the $I$ band light curve of \sdss\,, for
    $\Mwd=0.49$\,\msun\, and $\Msec=0.26$\,\msun. The model
    meets both the $\chi^{2}$ (within $1\sigma$) and the $\delta R$
    (within $15\%$) cut-offs. The residuals from the fit are shown at
    the bottom of the panel. Inset panel: data points and model fit
    focused around the eclipse phase.}
  \label{1212fit}
\end{figure}

Fig.\,\ref{1212fit} shows one example of the light curve fits within the
components masses range for the model parameters:
$\Mwd=0.49$ \msun, $\Msec=0.26$ \msun\,  and $i=89.2^{\circ}$. 
The detailed models do not predict any variation in the light curve caused by 
irradiation of the secondary star by the white dwarf. The predicted variations
due to ellipsoidal modulation are expected to be quite small, i.e. $\sim0.005$
mag, consistent with our observational non-detection of any variability
outside the eclipse. 

\subsubsection{Spectral energy distribution}
\label{sec:sed}
We cross-identified \sdss\, with the database from the Galaxy Evoluion
Explorer (GALEX \citep{martinetal05-1,morriseyetal05-1}), and found a detection in the far and near ultraviolet (FUV
and NUV). The magnitudes are $m_\mathrm{FUV} = 16.79 \pm 0.03$ mag and $m_\mathrm{NUV} = 16.81 \pm
0.02$ mag, exposure times were 150 sec.  FUV and NUV fluxes can provide an
estimate of the effective temperature of the white dwarf for a certain $\log
g$, assuming that all the flux in the UV is emitted by the primary. White
dwarf models for $\log g = 7.5$ and $\log g = 8.0$\, and effective temperature
in the range $6000-100000$ K, were folded over the FUV and NUV filters. The
calculated flux ratio FUV/NUV was compared with the observed for \sdss. The
GALEX flux ratio implies $\Teff\sim13000$ K, significantly colder than what we obtain from the optical spectrum in Sec.~\ref{fitting}. However, discrepant temperatures
from GALEX UV and optical photometry were noticed earlier from an analysis for a large
number of white dwarfs ($\sim250$) by \cite{2007astro.ph..2420K}. This shows
that one cannot expect the same UV and optical temperatures in a case-by-case
basis, but at best on a statistical average. For the time being we accept the
temperature from our fit to the SDSS spectrum, which grossly reflects the UV
to optical SED. 

\begin{figure}
\centering
\includegraphics[width=9.0cm,angle=0]{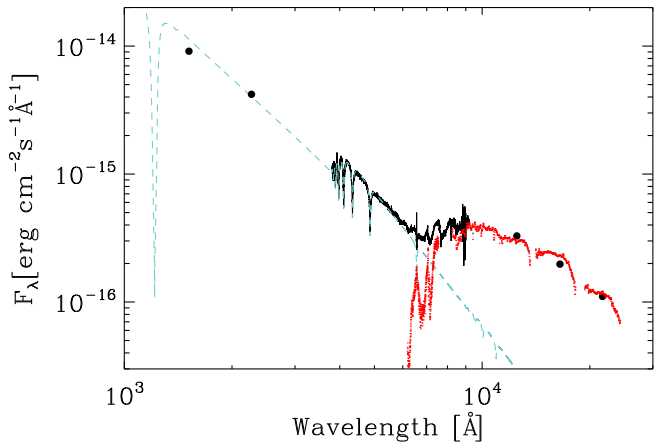}
\caption{Spectral energy distribution of \sdss. GALEX near and far
  ultraviolet and 2MASS infrared fluxes (black circles), optical SDSS
  spectrum (black line). A white dwarf model of $\Teff=17500$ K and
  $\log g=7.5$ (blue dashed line) and the spectum 
  of LHS1504 with spectral type $\mathrm{M5}$ from Legget's library (red 
  dots) are shown for comparison .}
 \label{g:all} 
\end{figure}

The spectral energy distribution is shown in Fig. \ref{g:all}, including
ultraviolet, optical and infrared fluxes from 2MASS. A model spectrum for a white dwarf of pure Hydrogen \citep{2005A&A...439..317K} with effective temperature of $17500$ K and $\log g = 7.5$ and a spectrum of the M5 star LHS1504 from Legget's library \footnote{http://ftp.jach.hawaii.edu/ukirt/skl/dM.spectra/} are shown for comparison \citep{Leggett2000}.

\subsubsection{Binary parameters summary}
\label{sum}

\begin{figure}
\centering
\includegraphics[width=10cm,angle=0,clip=]{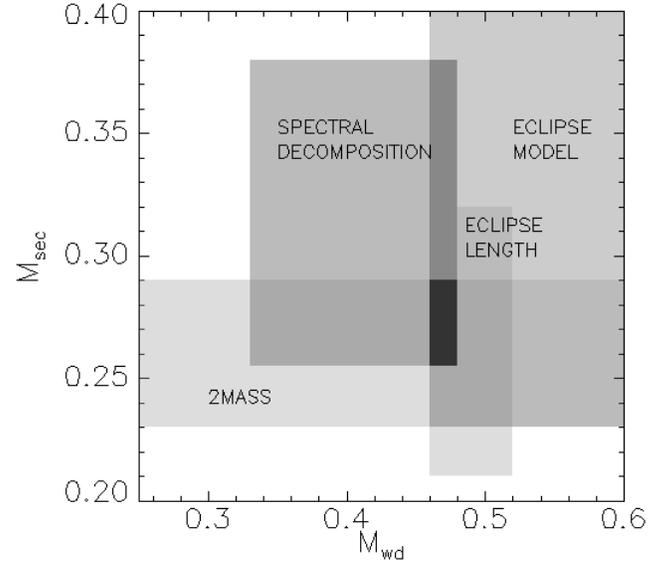}
\caption{The ranges of masses of the white dwarf and 
the red dwarf coming from: the decomposition of the SDSS spectrum; the infrared brightness; the eclipse length for an inclination of $90\degr$, and, the detailed light curve modelling.  
Each of the areas is labeled 
correspondingly, the intersection of the four different methods occurs 
for $\Msec = 0.26-0.29$ \msun\, and $\Mwd = 0.44-0.46$ \msun.} 
\label{g:masses}
\end{figure}

Figure \ref{g:masses} shows the different ranges for the masses of the primary and 
the secondary from the spectral decomposition fit (Sect.\,\ref{fitting}), the $K$-band
luminosity-mass relation (Sect.\,\ref{sec-2mas}), the radial 
velocity amplitude and eclipse length (Sect.\,\ref{period_rv}) and the detailed light 
curve fitting (Sect.\,\ref{sec-lcf}). Of course, the different methods are not entirely
independent, e.g. the constraints from the eclipse 
length/radial velocities studies and the detail light curve fitting basically use 
the same information with the only difference 
being that we could derive a clear lower limit from the latter.   
The dark shaded region in Fig.\,\ref{g:masses} 
represents the ranges of stellar masses in agreement 
with all the derived constraints i.e., $\Mwd = 0.46-0.48$ \msun,
$\Msec = 0.26-0.29$ \msun, implying a radius of the secondary star in the
range $\Rsec = 0.28-0.31$ \rsun\, using the empirical M--R relation from \cite{Bayless2006} 
and $\Rwd = 0.016-0.018$ \rsun\, ($\logg =7.5-7.7)$. We adopt these values
as the most probable ones and all finally accepted
stellar and binary parameters based on Sloan-data, other catalogues and our own
follow-up observations are collected in Table~\ref{t:param}.

\begin{table}[h]
\renewcommand{\footnoterule}{}
\caption{Stellar and binary parameters of \sdss.} %Coordinates, Sloan, GALEX and 2MASS magnitudes are quoted.
%$\Teff$, $\logg$, $\mathrm{Sp(2)}$ and $d$ are results from the deconvolution of the optical SDSS 
%spectrum. The ranges in the masses and radii are discussed in detail in Sec.\,\ref{sum}.
%>From the fit of the modeled $I$ band light curve a minimun inclination $i_\mathrm{min}$ is found.}
\centering
\label{t:param}       
\begin{tabular}{lrlr}
\hline\hline 
Parameter & Value & Parameter & Value\\
\hline\noalign{\smallskip}
$ R.A.\, (J2000.0)$  &  12 12 58.25             & $\Teff$ (K)           & $17700 \pm 300$     \\
$ Dec.\, (J2000.0) $ & -01 23 10.1              & $\logg$ (dex)         & $7.5-7.7$           \\
$ u $                & $17.045 \pm 0.020$       & $\mathrm{Sp(2)}$      & $\mathrm{M4} \pm 1$ \\
$ g $                & $16.769 \pm 0.013$       & $\Rwd$ (\rsun)        & $0.016-0.018$       \\
$ r $                & $16.936 \pm  0.013$      & $\Rsec$ (\rsun)       & $0.28-0.31$         \\
$ i $                & $16.627 \pm 0.015$       & $\Mwd$ (\msun)        & $0.46-0.48$         \\
$ z $                & $16.136 \pm 0.018$       & $\Msec$ (\msun)       & $0.26-0.29$         \\
$J $                 & $14.83 \pm 0.03$         & $P_{orb}$ (days)      & $0.3358706(5)$      \\
$H $                 & $14.35 \pm 0.05$         & $ K_2 $ (\kmps)       & $181 \pm 3$         \\
$K_s$                & $13.93 \pm 0.05$         & $ \gamma_2$ (\kmps)   & $17 \pm 3$          \\
$m_\mathrm{FUV}$     & $16.79 \pm 0.03$         & $\mathrm{a}$ (\rsun)  & $1.8 \pm 0.1$       \\
$m_\mathrm{NUV}$     & $16.81 \pm 0.02$         & $i_\mathrm{min}$      & $82\degr $          \\
$d$ (pc)             & $230 \pm 20 $            &                       &                     \\
\noalign {\smallskip} \hline
\label{t:param}
\end{tabular}
\end{table}

\section{Evolutionary state}
\label{evolu} 

The post CE evolution of compact binaries is driven by angular momentum loss
due to gravitational radiation and -- perhaps much stronger -- magnetic wind
braking. Unfortunately, the latter mechanism is currently far from being 
well constrained, and predicting and reconstructing the post CE evolution
sensitively depends on the assumed prescription for magnetic braking. 

However, the disrupted magnetic braking scenario proposed by 
\cite{rappaportetal83-1} can still be considered the standard model for
magnetic braking in close compact binaries. 
In this scenario it is assumed that magnetic braking ceases 
when the secondary star becomes fully convective at $\Msec\sim0.3$\msun\, (which corresponds to $\Porb\sim3$\,hrs). 
Although observations of the
spin down rates of single stars do drastically disagree with the predictions
of disrupted magnetic braking \citep{sillsetal00-1}, 
it remains the only consistent theory explaining the orbital period gap
i.e. the observed deficit of CVs in the range of
$\Porb\sim2-3$hrs. Moreover, first results of our radial velocity survey of
PCEBs seem to support the idea of disrupted magnetic braking 
\citep{Schreiber2008}. 
To predict and reconstruct the post CE evolution of \sdss\, according to 
\cite{Schr03}, we therefore assume disrupted magnetic braking. 

\begin{figure}[t!]
\centering
\includegraphics[width=6.75cm,angle=-90]{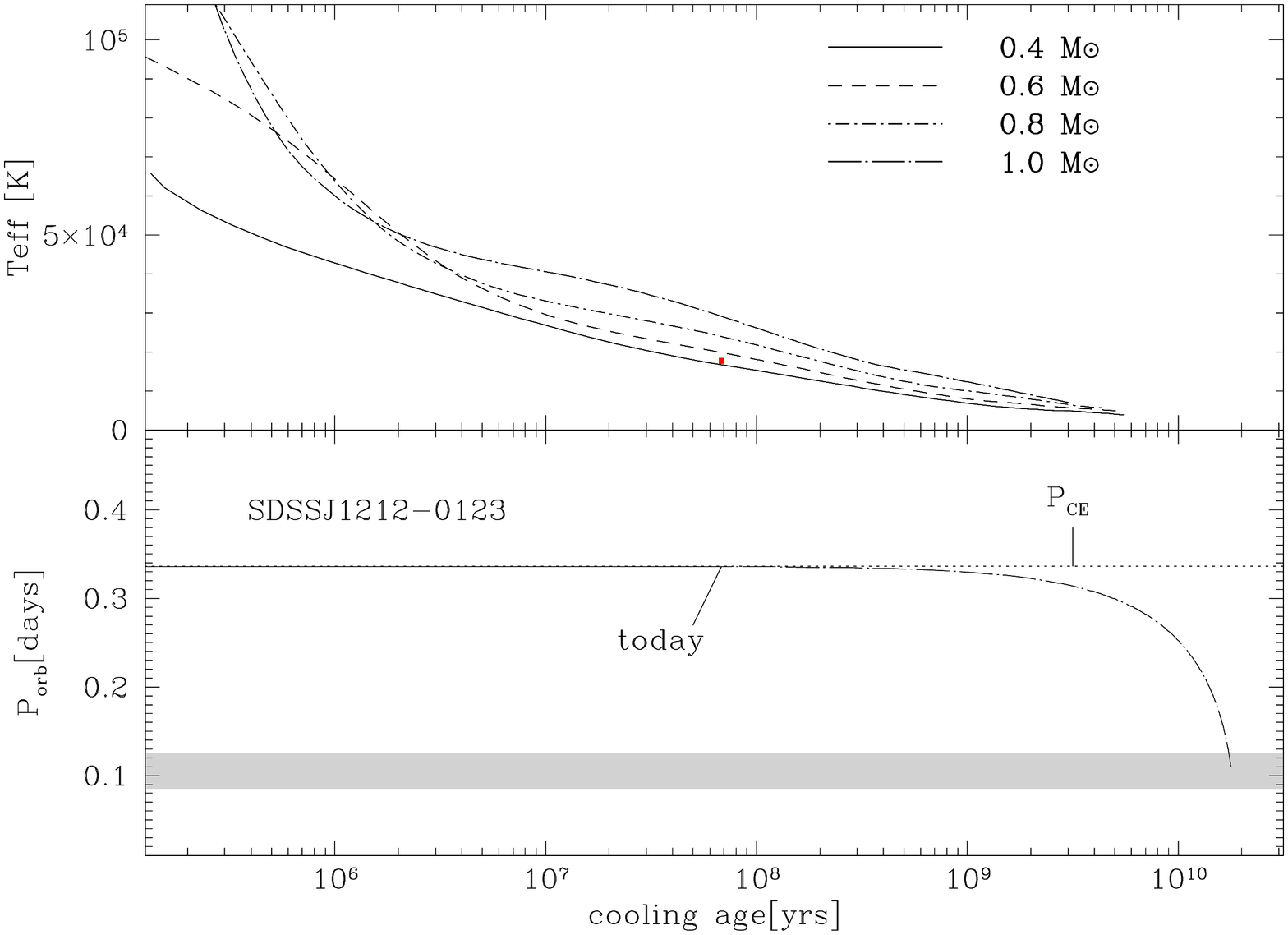}
\caption{Top panel: 
Interpolating the cooling tracks from \citet{wood95-1}
and according to the current temperature of the white dwarf ($\Teff=17700$\,K) 
we derive for \sdss\, a cooling age of $\sim7\times10^7$\,years. 
Bottom panel: Assuming gravitational radiation as
the only angular momentum loss mechanism we reconstructed the post-CE
evolution of \sdss\, and find that it left the CE phase with an 
orbital period of $P_{\mathrm{CE}}\sim\,8.07$\,hrs. Apparently, \sdss\, has passed only
a small fraction of its PCEB lifetime and it will take 
$\sim\,1.8\times\,10^{10}$ years until \sdss\, will become a CV. 
At that moment the white dwarf temperature
will be $\Teff\sim\,4000$\,K and the system will be inside the period gap (grey bar).}
\label{g:evol}
\end{figure}

First, we interpolate the cooling tracks of \cite{wood95-1} and estimate the cooling age of \sdss\, to be
$6.8\times10^7$\,yrs (see top panel of Fig.\,\ref{g:evol}).
Second, according to the mass 
derived for the secondary star ($\Msec\sim0.27$\msun) we assume that, 
since \sdss\, left the CE phase, the only mechanism driving the evolution 
of \sdss\, towards shorter orbital periods is (and has been) gravitational
radiation. 
As shown in Fig.\,\ref{g:evol} (bottom panel), \sdss\, 
left the CE phase with an orbital period of $P_{\mathrm{CE}}\sim\,8.07$\,hrs, 
very similar to the present value. 
Significant changes in the orbital period are predicted to occur on timescales
longer than the current cooling age of the white dwarf. 
In $\sim\,1.8\times\,10^{10}$ years \sdss\, will eventually become a
CV within the orbital period gap, however, giving that it's calculated PCEB 
lifetime exceeds the age of the Galaxy it is not representative of the 
progenitors of todays CV population.

\section{Conclusions}
\label{disc}
From optical photometry we conclude \sdss\, is a eclipsing PCEB with an
orbital period of $0.336$ days and an eclipse length of $23$ min. From
spectroscopic follow-up observations we have derived a systemic velocity of
$17\pm3 $ km/s and a semi-amplitude of the radial velocity of $181\pm3 $
km/s. From the SDSS spectrum we derived $\Teff = 17700 \pm 300
$ K, $\logg = 7.53 \pm 0.2$ implying a mass in the range $0.33-0.48$ \msun\, 
and a secondary spectral type $\mathrm{M4} \pm 1$, and a distance to the system of $230 \pm 20$ parsecs. 
From infrared photometry, using a mass--luminosity empirical relation we
derived $\Msec = 0.26 \pm 0.03$ \msun. 
We have calculated the radius of the secondary star using an empirical mass--radius ralation.
The mass function, combined with the eclipse length, points towards the high end
of the allowed mass range of the primary, i.e. $\Mwd\sim0.46-0.48$. 
We have modeled the $I$ band light curve and find 
the inclination of the orbit to be $i>82\degr$, and the masses to be
consistent with previously determined values. 
The different methods applied are all consistent with $\Mwd = 0.46-0.48$ \msun, implying 
$\Rwd = 0.016-0.018$ \rsun\, ($\logg=7.5-7.7$) for the primary and $\Msec =
0.26-0.29$ \msun, 
$\Rsec = 0.28-0.31$ \rsun\, for the 
secondary. We have reconstructed
and predicted the post CE evolution of \sdss, finding that \sdss\, 
at the end of the CE phase had a very similar orbital period. 
The only mechanism involved in
shrinking the orbital period is and has been gravitational radiation.
As the PCEB lifetime of \sdss\, exceeds the Hubble time we conclude that it 
is not representative of the progenitors of the current CV population. 

\begin{acknowledgements}
We thank our referee, Dr.S.B Howel, for a careful review of the original manuscript. We thank T.R. Marsh for the use of his light-curve modeling code. ANGM, MRS, 
RSC, JV and MK acknowledge support by the Deutsches Zentrum f\"ur Luft-und Raumfahrt (DLR) GmbH under contract No. FKZ 50 OR 0404. MRS was also supported by FONDECYT (grant 1061199), DIPUV (project\,35), and the Center of Astrophysics at the
Universidad de Valparaiso. JK was supported by the DFG priority programme SPP1177 (grant Schw536/23-1). 
    Funding for the SDSS and SDSS-II has been provided by the Alfred P. Sloan Foundation, the Participating Institutions, the National Science Foundation, the U.S. Department of Energy, the National Aeronautics and Space Administration, the Japanese Monbukagakusho, the Max Planck Society, and the Higher Education Funding Council for England. The SDSS Web Site is http://www.sdss.org/.   The SDSS is managed by the Astrophysical Research Consortium for the Participating Institutions. The Participating Institutions are the American Museum of Natural History, Astrophysical Institute Potsdam, University of Basel, University of Cambridge, Case Western Reserve University, University of Chicago, Drexel University, Fermilab, the Institute for Advanced Study, the Japan Participation Group, Johns Hopkins University, the Joint Institute for Nuclear Astrophysics, the Kavli Institute for Particle Astrophysics and Cosmology, the Korean Scientist Group, the Chinese Academy of Sciences (LAMOST), Los Alamos National Laboratory, the Max-Planck-Institute for Astronomy (MPIA), the Max-Planck-Institute for Astrophysics (MPA), New Mexico State University, Ohio State University, University of Pittsburgh, University of Portsmouth, Princeton University, the United States Naval Observatory, and the University of Washington. This publication makes use of data products from the Two Micron All Sky
Survey, which is a joint project of the University of Massachusetts and the
Infrared Processing and Analysis Center/California Institute of Technology,
funded by the National Aeronautics and Space Administration and the National
Science Foundation.
\end{acknowledgements}

\bibliographystyle{aa}
\bibliography{aabib}
\end{document}